\documentclass[aps,twocolumn,showpacs,superscriptaddress]{revtex4-1}
\usepackage{amsmath}
\usepackage{subfigure}
\usepackage{graphicx}
\usepackage{dcolumn}
\usepackage[colorlinks=true, citecolor=blue,linkcolor=blue,urlcolor=blue
]{hyperref}
\usepackage{color}
\usepackage{bm}

\newcommand{\beq}{\begin{equation}}
\newcommand{\eeq}{\end{equation}}

\begin{document}

\author{M. Asjad}
\affiliation{School of Science and Technology, Physics Division, University of Camerino, Camerino (MC), Italy}

\author{G. S. Agarwal}
\affiliation{Department of Physics, Oklahoma State University, Stillwater, Oklahoma 74078, USA}

\author{M. S. Kim}
\affiliation{QOLS, Blackett Laboratory, Imperial College London, SW7 2BW, United Kingdom}

\author{P. Tombesi}
\affiliation{School of Science and Technology, Physics Division, University of Camerino, Camerino (MC), Italy}

\author{G. Di Giuseppe}
\affiliation{School of Science and Technology, Physics Division, University of Camerino, Camerino (MC), Italy}

\author{D. Vitali}
\affiliation{School of Science and Technology, Physics Division, University of Camerino, Camerino (MC), Italy}

\title{Robust stationary mechanical squeezing in a kicked quadratic optomechanical system}


\begin{abstract}
We propose a scheme for the generation of a robust stationary squeezed state of a mechanical resonator in a quadratically coupled optomechanical system, driven by a pulsed laser. The intracavity photon number presents periodic intense peaks suddenly stiffening the effective harmonic potential felt by the mechanical resonator. These ``optical spring kicks'' tend to squeeze the resonator position, and due to the interplay with fluctuation-dissipation processes one can generate a stationary state with more than $13$ dB of squeezing even starting from moderately ``pre-cooled'' initial thermal states.
\end{abstract}

\pacs{42.50.Dv, 07.10.Cm, 42.50.Wk, 42.50.-p}
\maketitle

Squeezed states are characterized by an uncertainty of a single motional quadrature which is
below the zero-point level. Such states are particularly useful for ultrasensitive force detection~\cite{Caves1980}, as they are the optimal states of an harmonic oscillator providing the ultimate force sensitivity in the presence of losses~\cite{Latune2012}. Squeezed states are also important in quantum information, as they have been shown to represent a general resource for continuous variable quantum information processing~\cite{Braunstein2005}. Coherent parametric driving is the most popular and direct scheme for generating squeezing of a bosonic mode, and for a mechanical resonator it is realized
by modulating the spring constant at twice the mechanical resonance frequency~\cite{Walls1995}. However, it can achieve at best steady-state squeezing
$50\%$ below the zero-point level (the so-called 3 dB limit)~\cite{Walls1995}, because by further increasing the modulation strength, the system becomes unstable.

The recent rapid development of cavity optomechanics~\cite{Kippenberg2007,Marquardt2009,Genes2009} describing the interaction and mutual control between mechanical and cavity electromagnetic modes, has provided new paths and opportunities for the generation and manipulation of squeezing. Quadrature squeezing of the cavity output light due to the optomechanical interaction has been recently demonstrated~\cite{Safavi-Naeini2013,Purdy2013a}, almost $20$ years after its prediction~\cite{Mancini1994,Fabre1994}. Mechanical squeezing has not been achieved yet, even though various schemes for its generation have been proposed. Coherent parametric driving is easily achieved by modulating the radiation pressure force as in~\cite{Woolley2008,Mari2009,Nunnenkamp2010,Liao2011,Schmidt2012}. Better squeezing, below the 3 dB limit, can be obtained by continuously injecting squeezed light directly
into the cavity~\cite{Jahne2009}, but this is difficult as one needs a source of highly squeezed light and a high state-transfer efficiency at the quantum level. The 3 dB limit can also be beaten with closed-loop controls, i.e., exploiting continuous quantum feedback~\cite{Clerk2008,Szorkovszky2011,Szorkovszky2013,Pontin2013,Vinante2013}, but they are not easy to implement in the quantum regime, as they are seriously limited by non-unit detection efficiency and typically require fast measurements and feedback loops.

Here we show that robust \emph{stationary} mechanical squeezing, more than $13$ dB below the vacuum level, is achievable in a quadratic optomechanical system by exploiting \emph{impulsive open-loop} controls, which are realized by driving the cavity with a suitably pulsed laser. The intracavity photon number is periodically peaked, correspondingly stiffening the harmonic potential, and realizing ``optical spring kicks''. Such an impulsive ``bang-bang'' scheme is related with dynamical decoupling schemes~\cite{Vitali1999,Viola1999,Vitali2001a} used for decoupling systems from their environment and suppressing decoherence. Here, however, optical spring kicks rather \emph{cooperate} with environmental fluctuation-dissipation processes, selecting and stabilizing phase-dependent nonclassical fluctuations associated with the stationary squeezed state. This scheme differs from those of Refs.~\cite{Kronwald2013,Didier2013}, exploiting reservoir engineering schemes~\cite{Poyatos1996,Carvalho2001}, and which require operation in the well-resolved sideband regime. These latter schemes are limited by the effect of counter-rotating terms in the optomechanical interaction, while the present scheme is instead optimal in the unresolved sideband regime $\kappa \gg \omega_m$.

We consider an optomechanical system formed by a driven cavity mode interacting \emph{quadratically} with a mechanical resonator (MR). Such a quadratic interaction is achieved in a membrane-in-the-middle (MIM) setup, when the membrane is placed at a node, or exactly at an avoided crossing point within the cavity \cite{Thompson2008,Sankey2010,Karuza2013,Flowers-Jacobs2012}. Alternatively, one can consider levitating nanoparticles trapped around an intensity maximum of a cavity mode~\cite{Barker2010,Li2011,Gieseler2012,Kiesel2013}.
We assume that the cavity mode is resonantly driven, that is
\begin{eqnarray}
&&H=\frac{\hbar \omega_m}{2}(p^2+q^2) +\hbar \omega_c a^{\dagger }a +\hbar g_2 a^{\dagger }a q^2\\
&&+\mathrm{i}\hbar \left[E_0(t) e^{-\mathrm{i}\omega_c t}a^{\dagger }-E_0 (t) e^{\mathrm{i}\omega_c t}a\right],  \nonumber \label{eq:Ham-optomech}
\end{eqnarray}
where $\omega_m$ is the resonance frequency of the MR, $q$ and $p$ its dimensionless position and momentum operators such that $[q,p]=i$, $a$ is the cavity mode annihilation operator and $\omega_c$ its frequency, $E_0(t)=\sqrt{2P_{0}(t)\kappa_0 /\hbar \, \omega_{c}}$, with $\kappa_0$ the cavity decay rate through the input mirror and $P_0(t)$ is the time-dependent input power. Finally $g_2$ is the quadratic optomechanical coupling rate.

The cavity is intensely driven so that the intracavity field is well described in terms of its classical amplitude $\alpha(t) = \langle a(t)\rangle$, satisfying the evolution equation $ \dot{\alpha}=-\kappa \alpha-i g_2 q^2 \alpha +E_0(t)$ in the frame rotating at the laser frequency,
where $\kappa=\kappa_0+\kappa_L$ is the total decay rate, with $\kappa_L$ the rate associated with photon losses due to transmission through the other mirror, absorption and scattering.
The quadratic coupling is typically small, so that $\kappa \gg g_2 \langle q^2 \rangle $ and we can neglect the effect of the MR on the intracavity amplitude, which is given by the very small fluctuating detuning caused by the MR position fluctuations. Therefore, we have the following intracavity field amplitude
    $ \alpha(t)= \alpha(0)e^{-\kappa t}+\int_0^{t} ds E_0(s) e^{-\kappa (t-s)}$,
and we end up with the purely mechanical Hamiltonian,
\begin{eqnarray}
H=\frac{\hbar \omega_m}{2}p^2  +\frac{\hbar}{2}\left[\omega_m+ 2 g_2 |\alpha(t)|^2\right] q^2 ,\label{eq:Ham-optomech4}
\end{eqnarray}
describing a resonator with a spring constant whose time-dependence can be controlled by the driving laser.

The MR is unavoidably coupled to its thermal reservoir, causing fluctuation-dissipation processes. Its effect can be described by means of quantum Langevin equations (QLE), obtained by adding to the Heisenberg equations of motion associated with Eq.~(\ref{eq:Ham-optomech4}), damping with rate $\gamma_m$ and a noise term~\cite{Giovannetti2001}
\begin{subequations}
\label{nonlinlang}
\begin{eqnarray}
\dot{q}&=&\omega_m p, \\  \dot{p}&=&-\left[\omega_m +2 g_2 |\alpha(t)|^2\right] q - \gamma_m p + \xi,
\end{eqnarray}
\end{subequations}
where $\xi$ is Gaussian quantum stochastic force with zero mean value and correlation function
\begin{equation}\label{browncorre}
\left \langle \xi(t) \xi(t')\right \rangle = \frac{\gamma_m}{\omega_m} \int_0^{\Omega_c}
  \frac{d\omega}{2\pi} e^{-i\omega(t-t')} \omega \left[\coth\left(\frac{\hbar \omega}{2k_BT}\right)+1\right]
\end{equation}
($k_B$ is the Boltzmann constant, $\Omega_c$ is the frequency cutoff of the reservoir, and $T$ is the temperature of the membrane).

We now assume that the driving laser is pulsed, with pulses of duration $\tau_{p}$ and separated by a time interval $\tau$. A satisfactory control of $|\alpha(t)|^2$ is obtained only if the pulsed laser does not drive other nearby cavity modes. In a MIM this is achieved by placing the membrane at a field node, by using a spatial mode cleaner so that transverse modes are not excited, and if the pulse bandwidth is smaller than the cavity free spectral range, $1/\tau_{p} < c/2L$ ($L$ is the cavity length). We also assume that the cavity decay time is much longer than the driving pulse duration but much smaller than the pulse separation, that is, $1/\tau_{p} \gg \kappa \gg 1/\tau $. In this way the cavity mode is quickly excited and then decays to the vacuum state before the arrival of the next pulse. Therefore, also the intracavity photon number $|\alpha(t)|^2$ has a pulsed time dependence with period $\tau$, and is nonzero only within a short time interval $\Delta t$ of the order of $1/\kappa$. The periodically pulsed laser therefore realizes a MR subject to \emph{optical spring kicks}, which tend to squeeze its position variance through the periodic strong increase of the confining harmonic potential (see~\cite{Agarwal2011} for a first description of the scheme). We now show that such a pulsed optomechanical scheme is able to generate a stationary squeezed state of the MR even starting from moderately cooled systems and using state-of-the-art apparatuses.

We are interested in the long-time dynamics of the MR, and since we have assumed $\kappa \tau \gg 1$, we can safely approximate the effect of the pulsed time dependence of the intracavity photon number in terms of a Dirac delta, that is, $|\alpha(t)|^2 \propto \delta(t)$. From Eq.~(\ref{eq:Ham-optomech4}), one has that a single kick is described by the unitary operator
\begin{equation}\label{eq:kick}
    U_K=e^{i \theta q^2},
\end{equation}
where $\theta = 2 g_2\int_{\Delta t} dt |\alpha(t)|^2$ is the dimensionless parameter quantifying the effect of the kick. It is immediate to see that the kick performs the transformation $q \to q$ and $p \to p-2\theta q$.

Between the above kicks the cavity is empty (at optical frequencies we can neglect thermal photons) and therefore for a time interval of duration $\tau$ the MR is subject to the dissipative and noisy dynamics described by Eqs.~(\ref{nonlinlang}) with $\alpha(t)=0$.
Eq.~(\ref{browncorre}) shows that $\xi(t)$ is generally a non-Markovian Gaussian stochastic process with a nonzero correlation time; however Eqs.~(\ref{nonlinlang}) reduce to a Markovian dynamics within the time interval $\tau$ at large enough frequency cutoffs and temperatures, $\Omega_c \tau \gtrsim 1$ and $k_B T \tau/\hbar \gtrsim 1$, where $\xi(t)$ becomes delta-correlated~\cite{Giovannetti2001,Gardiner2000},
\begin{equation}\label{browncorre2}
\left \langle \xi(t) \xi(t')\right \rangle \simeq (2\bar{n}+1)\gamma_m\delta(t-t')+i \frac{\gamma_{m}}{\omega_m}\delta'(t-t'),
\end{equation}
where $\bar{n}=\left(\exp\{\hbar \omega_m/k_BT\}-1\right)^{-1} \simeq k_B T/\hbar \omega_m -1/2$ is mean thermal phonon number, and $\delta'(t-t')$ is the derivative of the Dirac delta. The condition $\Omega_c \tau \gtrsim 1$ is easily satisfied because typical systems have $\Omega_c \gg \kappa$ and we have already assumed $\kappa \tau \gg 1$. The condition $k_B T \tau/\hbar  \simeq \bar{n} \omega_m \tau \gtrsim 1$ is more stringent, but is satisfied under current experimental conditions.

If the MR starts from a Gaussian state, both the kicks and the linear dissipative evolution between them preserve the Gaussian nature of the MR state. In particular it is interesting to consider an initial thermal state with equilibrium thermal phonon number $\bar{n}$, which is a Gaussian state with zero first order moments. In this case the mean values $\langle q \rangle$ and $\langle p \rangle$ remain equal to zero at all times, and the dynamics is fully described by the time evolution of the second order moments $\sigma_q=\langle q^2 \rangle$, $\sigma_{qp}=\langle qp+pq \rangle/2$ and $\sigma_p=\langle p^2 \rangle$.
The corresponding equations of motion are obtained using Eqs.~(\ref{nonlinlang}) and Eq.~(\ref{browncorre2}) and are given by~\cite{Gardiner2000}
\begin{subequations}
\label{eq:variance}
\begin{eqnarray}
    \dot{\sigma}_q &=& 2\omega_m \sigma_{qp}, \\
    \dot{\sigma}_{qp} &=& \omega_m (\sigma_p- \sigma_q) -\gamma_m \sigma_{qp},\label{eq:variance2}\\
    \dot{\sigma}_p &=& -2\omega_m \sigma_{qp} -2\gamma_m \sigma_p +\gamma_m(2\bar{n}+1). \label{eq:variance3}
\end{eqnarray}
\end{subequations}
Using the three-dimensional vector $v(t)=(\sigma_{q}(t),\sigma_{qp}(t),\sigma_{p}(t))^T$, the solution of these equations can be written in compact form as
$$v(t)=M(t) v(0)+v_{\rm inh}(t),$$ where $M(t)$ is a $3\times 3$ matrix which decays to zero at large $t$ due to damping and $v_{\rm inh}(t)$ is an inhomogeneous vector term which is proportional to the thermal equilibrium values $\bar{n}+1/2$. $M(t)$ and $v_{\rm inh}(t)$ can be easily calculated from Eqs.~(\ref{eq:variance}). The effect of a kick on the second order moments is instead given by
\begin{equation}\label{eq:kick2}
    v'=Kv \;\;\;\ K=\left(
\begin{array}
[c]{ccc}%
1 & 0 & 0\\
-2 \theta & 1 & 0 \\
4 \theta ^2 & -4 \theta &  1
\end{array}
\right).
\end{equation}
Therefore one has a cyclic evolution of duration $\tau$ which is described by the map
\begin{equation}\label{eq:iter}
v\left[(n+1)\tau\right]=M(\tau)K v\left[n\tau\right]+v_{\rm inh}(\tau).
\end{equation}
Iteration of this formula gives the solution of the stroboscopic dynamics of the MR at times $n \tau$
\begin{equation}\label{eq:map}
  v\left[n\tau\right]= A(\tau,\theta)^n v(0)+\frac{I-A(\tau,\theta)^n}{I-A(\tau,\theta)}v_{\rm inh}(\tau),
\end{equation}
with the $3\times3$ matrix $A(\tau,\theta)=M(\tau)K$. Since $\lim_{n \to \infty} A(\tau,\theta)^n =0$, the stroboscopic dynamics tends to a stationary Gaussian state of the MR characterized by the following second order moments
   $ v(\infty)=\lim_{n\to \infty} v\left[n\tau\right]=\left[I-A(\tau,\theta)\right]^{-1}v_{\rm inh}(\tau)$.

The MR state is squeezed when there is a mechanical quadrature $q(\varphi)=q\cos \varphi +p \sin \varphi$ with variance below $1/2$, which is the vacuum noise level in our definitions, i.e. if there is a phase $\varphi$ such that $\langle q(\varphi)^2\rangle < 1/2$. By minimizing with respect to $\varphi$, one has that
\begin{equation}\label{eq:varmin}
    \sigma_{\rm min} = \min_{\varphi} \langle q(\varphi)^2\rangle= \frac{1}{2}\left[\sigma_p+\sigma_q-\sqrt{(\sigma_p-\sigma_q)^2+4\sigma_{qp}^2}\right].
\end{equation}
We expect intuitively to achieve better squeezing for increasing kick strength $\theta$ and for decreasing time separation $\tau$, because more frequent kicks  make harder for the MR to reach the thermal equilibrium values $\sigma_q^{\rm th}=\sigma_q^{\rm th}=\bar{n}+1/2$. We now show that significant stationary mechanical squeezing can be achieved by driving with appropriate pulses a state-of-the-art membrane-in-the-middle setup like that of Ref.~\cite{Flowers-Jacobs2012}. In fact, choosing a cavity with $L=0.1$ mm and finesse $F \simeq 40000$, one has $\kappa \simeq 10^{8}$ s$^{-1}$, and assuming a driving laser with wavelength $\lambda = 1550$ nm, and pulses of duration $\tau_{p} \simeq 0.1$ ns, peak power $P \simeq 1$ W, separated by $\tau = 10^{-7}$ s, the above assumptions $c/2L > \tau_{p}^{-1} \gg \kappa \gg \tau^{-1}$ are satisfied, and the intracavity photon number achieves peak values $|\alpha(t)|^2 \simeq 6.7 \times 10^{10}$. Placing the membrane (with mass $m$ and reflectivity $R$) at a node one has $g_2 =\left(16 \pi^2 c \hbar/\lambda^2 L m \omega_m\right)\sqrt{R/(1-R)}$,
so that choosing $m=0.25 \times 10^{-11}$ kg, $R\simeq 0.2$, and $\omega_m=0.5\times 10^6$ s$^{-1}$, one gets $g_2 \simeq 0.8 \times 10^{-2}$ s$^{-1}$ and $\theta \simeq 10$.

\begin{figure}
\centering
\includegraphics[width=.49\textwidth]{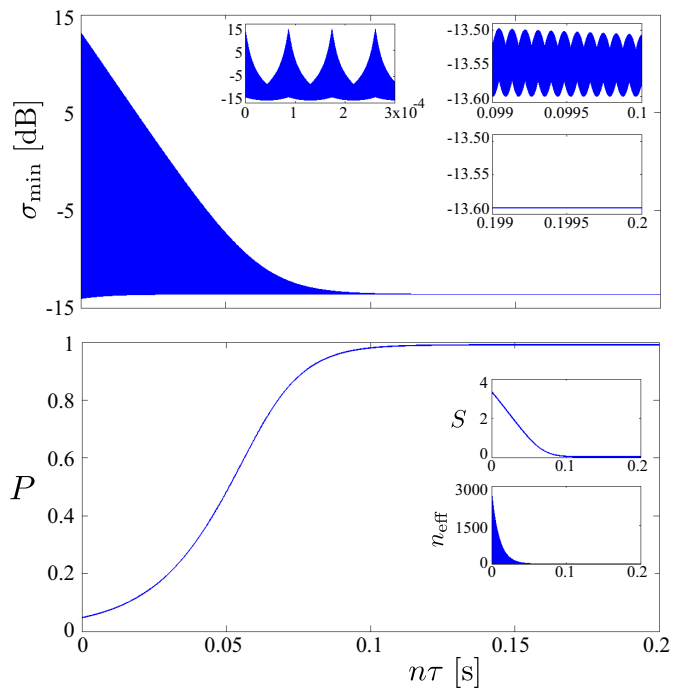}
\caption{(a) The minimum quadrature variance $\sigma_{\rm min}$ [in dB with respect to the vacuum level, i.e., $10\log_{10}\left(2\sigma_{min}\right)$] versus $n \tau$ for $\bar{n}=10$. The insets show a fine-grained view of the stroboscopic dynamics at different times. (b) Purity (and in the insets, von Neumann entropy $S$ and effective occupancy $n_{\rm eff}$) of the MR state versus $n\tau$. The other parameters are $\omega_m=0.5 \times 10^6$ s$^{-1}$, $\gamma_m=10^2$ s$^{-1}$, $\tau=10^{-7}$ s, $\theta = 10$.} \label{evo1}
\end{figure}

\begin{figure}
\centering
\includegraphics[width=.49\textwidth]{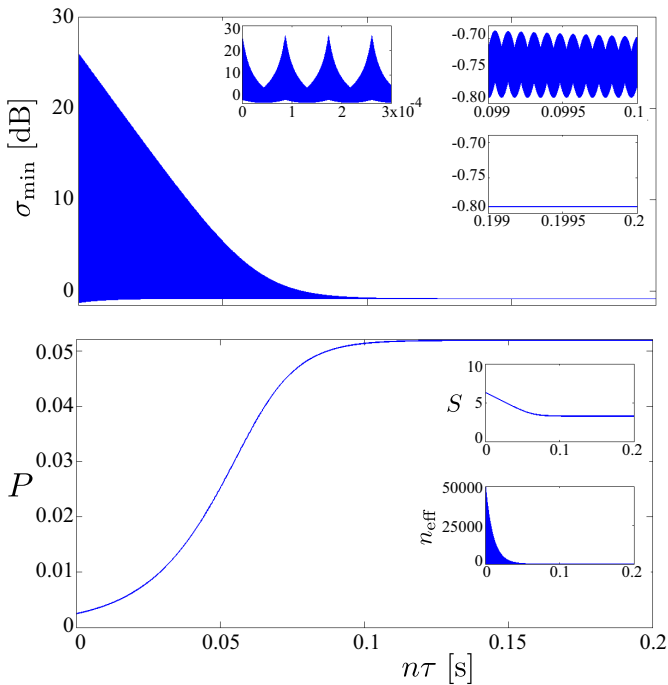}
\caption{The same as in Fig.~\protect\ref{evo1}, except that $\bar{n}=200$. One still get squeezing, but the steady state if far being pure.}
\label{evo3}
\end{figure}

We show the stroboscopic evolution of the mechanical state in the case of damping $\gamma_m=10^2$ s$^{-1}$ in Figs.~\ref{evo1}-\ref{evo3}. We plot the minimum variance $\sigma_{\rm min}$ (top) and the purity of the MR Gaussian state $\rho(n\tau)$, $P={\rm Tr}\left[\rho(n\tau)^2\right]=\left[4\left(\sigma_p \sigma_q-\sigma_{qp}^2\right)\right]^{-1/2}$ (bottom) versus $n \tau$ for $\bar{n}=10$ in Fig.~\ref{evo1} and $\bar{n}=200$ in Fig.~\ref{evo3}. We recall that $\bar{n}$ determines both the initial thermal state of the MR and its dynamics between the kicks according to Eqs.~(\ref{eq:variance}).

When $\bar{n}=10$, the MR stroboscopically tends to a strongly squeezed stationary state, more than 13 dB below the vacuum level; moreover, stationary mechanical squeezing of $0.8$ dB is achieved even when starting from $\bar{n}=200$. Optical spring kicks also \emph{purify} the MR state, and in particular when $\bar{n}=10$, the steady state is practically a pure minimal uncertainty squeezed state. This is confirmed by the decay to zero of the von Neumann entropy of the state $S$, which is related to the purity $P$ by $S=\left[(1-P)/2P\right]\log\left[(1+P)/(1-P)\right]-\log\left[2P/(1+P)\right]$~\cite{Agarwal1971} [see the inset in Fig.~\ref{evo1}(b)]. The purification provided by the optical spring kicks also asymptotically ``cools'' the MR, as shown by the decay of the effective mechanical occupancy $n_{\rm eff}=(\sigma_p +\sigma_q-1)/2$ [see the inset in Figs.~\ref{evo1}(b)-\ref{evo3}(b)]. In the steady state one has approximately position squeezing because the phase with minimum variance is $\varphi_{\rm min} \simeq -\omega_m \tau/8 \simeq -0.006$ so that $\sigma_{\rm min} \simeq \sigma_q$. The stationary squeezed state is reached with fast oscillations associated to the frequency $\omega_m$ [see the zoomed insets in Figs.~\ref{evo1}(a)-\ref{evo3}(a)], and squeezing below the vacuum noise level is steadily achieved after about $10^5$ kicks.

\begin{figure}
\centering
\includegraphics[width=.49\textwidth]{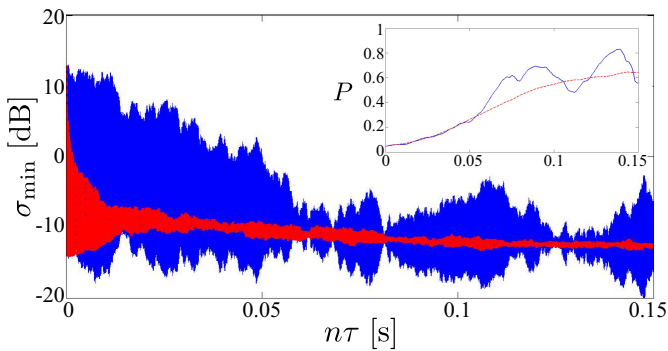}
\caption{The minimum quadrature variance $\sigma_{\rm min}$ (in dB with respect to the vacuum level), and the purity $P$ (in the inset) versus $n \tau$ in the case when $\theta$ randomly fluctuates from kick to kick according to a Gaussian distribution centered around $\theta=10$ and with variance 0.001. The other parameters are as in Fig.~\protect\ref{evo1}. The blue full line refers to a single experimental run, while the red dotted line refers to the average over $100$ trajectories.}
\label{rnd}
\end{figure}

In Fig.~\ref{rnd} we show that the proposed scheme for generating stationary mechanical squeezing is robust with respect to fluctuations of the laser pulses. In fact Fig.~\ref{rnd} shows the stroboscopic time evolution of $\sigma_{\rm min}$ (a) and of the purity $P$ (b) in the same set of parameters of Fig.~\ref{evo1}, except that now the kick parameter $\theta$ randomly changes from kick to kick according to a Gaussian distribution centered around $\theta =10$ and with variance equal $0.001$, modeling pulse area fluctuations at $0.3\%$ level. The blue full line refers to a single experimental run, while the red dotted line refers to the average over $100$ trajectories. One gets large squeezing even in the presence of appreciable fluctuations,

Reaching the chosen values $\bar{n} \leq 200$ only with cryogenic techniques is very hard, since it requires operating below 1 mK. However they could be achieved with a standard laser cooling scheme~\cite{Teufel2011a,Chan2011}. Therefore the whole experiment could be realized by driving a high-finesse cavity with two lasers at well distinct wavelengths: a first laser drives a mode linearly interacting with the membrane and provides cooling; the second one is the pulsed laser considered above, which interacts quadratically with the membrane. The quite large value of mechanical damping $\gamma_m=10^2$ s$^{-1}$ considered here is consistent with the presence of an additional moderate laser cooling process.

One could detect the generated squeezed state using the scheme suggested in Ref.~\cite{Vitali2007} or the pulsed homodyne measurement scheme of Ref.~\cite{Vanner2011}, both requiring an additional probe field linearly interacting with the MR.  However, one can exploit again the quadratic optomechanical interaction for a \emph{direct} detection of squeezing. In fact, in the bad cavity limit $\kappa \gg \omega_m $ we are considering, and at first order in $g_2$, one has that the output phase quadrature $Y_{\rm out}(t)=-i[a_{\rm out}(t)-a_{\rm out}(t)]$ of any probe field quadratically interacting with the MR is given by
\begin{equation}
Y_{\rm out}(t) \simeq Y_{\rm out}^{0}(t)+\frac{g_2}{\kappa}q^2(t)X_{\rm out}^{0}(t),
\end{equation}
where $X_{\rm out}^{0}(t)=a_{\rm out}^0(t)+a_{\rm out}^0(t)$ is the output amplitude quadrature of the probe and the index $0$ denotes the zeroth order output field without the optomechanical interaction. Therefore a homodyne measurement of the probe allows to detect directly the squeezed position variance of the MR.

In conclusion, we have shown an open loop ``bang bang'' control which could generate \emph{stationary} mechanical squeezing more than $13$ dB below the vacuum noise level, using state-of-the-art apparata and starting from moderately cooled mechanical states. Stationary squeezing is obtained by using optical spring kicks, that is, by appropriately pulsing an optomechanical cavity with quadratic interaction. Such a robust and large mechanical squeezing could provide unprecedented force sensitivity~\cite{Latune2012}.

\emph{Acknowledgments}
This work has been supported by the European Commission (ITN-Marie Curie project cQOM), and by MIUR (PRIN 2011).

\bibliography{cat-state1}

\end{document}